\documentclass[a4paper,11pt]{article}

\usepackage{jcappub}
\usepackage[T1]{fontenc}
\usepackage[normalem]{ulem}
\usepackage{amsmath}
\usepackage{amssymb}
\usepackage{geometry}
\usepackage{booktabs}
\usepackage{array}
\usepackage{paralist}
\usepackage{verbatim}
\usepackage{graphicx}
\usepackage[utf8]{inputenc}
\usepackage{multirow}
\usepackage[dvipsnames*,svgnames]{xcolor}
\usepackage{wasysym}
\usepackage{subfigure}
\usepackage{enumitem}
\usepackage{lineno}
\usepackage{float}

\title{\textbf{Fermi acceleration under Lorentz invariance violation}}
\author{Matheus Duarte and}
\emailAdd{matheus\_duarte@usp.br}

\author{Vitor de Souza}
\emailAdd{vitor@ifsc.usp.br}

\affiliation{Instituto de F\'isica de S\~ao Carlos, Universidade de S\~ao Paulo, Av. Trabalhador S\~ao-carlense 400, S\~ao Carlos, Brasil}

\abstract{In this paper, the acceleration of particles in astrophysical sources by the Fermi mechanism is revisited under the assumption of Lorentz invariance violation (LIV). We calculate the energy spectrum and the acceleration time of particles leaving the source as a function of the energy beyond which the Lorentz invariance violation becomes relevant. Lorentz invariance violation causes significant changes in the acceleration of particles by the first and second-order Fermi mechanisms. The energy spectrum of particles accelerated by first-order Fermi mechanism under LIV assumption shows a strong suppression for energies above the break. The calculations presented here complete the scenario for LIV searches with astroparticles by showing, for the first time, how the benchmark acceleration mechanisms (Fermi) are modified under LIV assumption.}

\begin{document}

\maketitle
\flushbottom

\section{Introduction}

The principle of relativity and Lorentz invariance (LI) are the fundamental pillars of quantum field theory and, therefore, the basis for our understanding of Nature~\cite{Einstein:1905ve}. Several tests of these principles were done~\cite{Ives:38,PhysRev.42.400}, and no departure was found. Despite that, general relativity and quantum theories remain irreconcilable. Various unification formalisms were proposed, including string theory~\cite{veneziano,GREEN1981502}, loop quantum gravity~\cite{Polyakov:1987zb} and curved momentum space~\cite{Doplicher_1995} among others; nonetheless, a truly compelling and robust unification theory remains elusive. Lorentz invariance violation (LIV) is a hypothesis that could pave the way for the development of such a model~\cite{Greenberg_2002,Mattingly_2005}. Amidst the difficulty of finding experiments to validate or disprove quantum gravity theories, the pursuit of deviations from Lorentz invariance assumes critical importance as it holds the potential to either invalidate or refine a broad spectrum of models.

Ultra-high energy cosmic rays (UHECRs) ($E> 10^{17}$ eV) are the most energetic particles known in the Universe and, therefore, a promising probe for LIV. From the creation to the detector at Earth, these particles travel through very different media: a) the source environment, b) the extragalactic medium and c) the atmosphere. Several searches for LIV using UHECR were done, and stringent limits on the invariance parameters were set~\cite{Amelino-Camelia:1997ieq,Jacob_2008,Zhang_2015,Schreck_2017,Abreu_2022,Kosteleck__2015,Guedes_Lang_2018,Albert_2020,Satunin_2019,Duenkel_2021} by studying the propagation in the extragalactic medium and the development of the cascade of particles in the Earth's atmosphere. Still, it is unknown to us whether any study of Lorentz invariance violation by UHECRs in the source was performed. If UHECRs break LI, they will do it first in the acceleration procedure in the source.

The sources of UHECR are yet to be identified~\cite{Anchordoqui_2019}. The most promising candidates are those able to keep Fermi acceleration in operation, such as extragalactic active galactic nuclei (AGN), radio-galaxies and starburst regions. In 1949, Fermi proposed an acceleration model for UHECR based on interactions of the particles with shock waves~\cite{PhysRev.75.1169}. This original model was improved~\cite{Vietri_2003,Blasi_2005}, and two branches called first and second-order Fermi acceleration were developed. In both cases, Lorentz invariance is assumed, and it plays a decisive role in determining the maximum energy achievable by a particle in each source, the energy spectrum produced by the source, and its luminosity.

In this paper, we investigate the Fermi acceleration mechanisms of UHECRs under the hypothesis of LIV. In section~\ref{sec:liv}, we define the LIV framework used. In sections~\ref{sec:second:order} and~\ref{sec:first:order}, we deduce the maximum energy, the energy spectrum and acceleration time of UHECR accelerated by Fermi's second and first-order mechanisms, respectively, under the LIV hypothesis. In section~\ref{sec:conclusion}, we conclude the work.

\section{Lorentz invariance framework}
\label{sec:liv}

From a phenomenological viewpoint, LIV can be expressed in special relativity by a modification of Einstein's dispersion relation introducing a function $f(p,m)$ as
\begin{equation}
    E^2 = m^2 + p^2 + f(p,m) \textrm{,}
\end{equation}
where we have used $c=1$. At energies accessible in laboratories, experiments show $f(p,m) \rightarrow 0$, allowing to expand $f(p,m)$ in terms of $p$ as
\begin{equation}
    E^2 = m^2 + p^2 + \sum_{n} \delta_n p^{n+2} \textrm{,}
\end{equation}
where $\delta_n$ are small breaking factors in comparison to the energy scale in which the LIV is supposed to be relevant. Experiments have set strong limits on $\delta_n$ for energies up to $10^{20}$ eV~\cite{Galaverni_2008,sym12081232}. From a theoretical point of view, LIV is expected to be relevant at the Planck scale,  $E_{pl} \approx 10^{28}$ eV~\cite{Pfeifer:2023cgd}. In any case, we can safely assume $p \gg m$, leading to
\begin{equation}
  p = \frac{E}{\sqrt{ 1 +  \delta_n E^{n}}} \textrm{.}
  \label{eq:liv:edisper}
\end{equation}

This equation summarizes LIV in the following calculations for the Fermi acceleration.

%==================================================================
\section{Second-order Fermi acceleration}
\label{sec:second:order}

The original argument elaborated by Fermi is nowadays called second-order Fermi acceleration~\cite{PhysRev.75.1169}. It is based on collisions of particles with moving gas clouds carrying magnetized plasma. Below, we review Fermi's calculations of the energy gain of an ensemble of particles going through collisions, including the LIV assumption in the original argument. Two reference frames are used for the calculations: the cloud and the laboratory defined as the reference frame in which the cloud has velocity $V$. Given that the mass of the cloud is much larger than the mass of the particles, the velocity of the cloud $V$ does not change with the interaction and, therefore, the cloud reference frame is the center of momentum frame.

A particle with energy $E$ and momentum $p$ measured in the laboratory reference frame has energy $E'$ in the cloud reference frame
\begin{equation}
    E' = \gamma_V (E + Vp\cos{\theta}) \textrm{,}
\end{equation}
where $\theta$ is the angle between the cloud's velocity and the particle's velocity.

In the cloud frame, the collision reverses the momentum of the particle. Taking that into account, we obtain the energy of the particle after the collision $E''$ in the laboratory reference frame
\begin{equation}
    E'' = \gamma_V (E' + Vp_x') \textrm{.}
\end{equation}
Using equation~\ref{eq:liv:edisper} of the energy dispersion relation modified by LIV, we can write the energy gain of the particle after the collision as
\begin{equation}
    \frac{E''- E}{E} = \gamma_V^2 \left(1 + \frac{2V\cos{\theta}}{\sqrt{1 + \delta_n E^n}} + V^2\right) - 1 \textrm{.}
\end{equation}
If $\delta_n=0$, the original result from Fermi is obtained.

For an ensemble of particles, the average energy gain can be calculated by considering that the chance of collision at an angle $\theta$ is proportional to $\gamma_V\left(1 + V\cos{\theta}\right)$~\cite{Longair_2011}. Computing the average of the cosine and taking the terms up to second order in $V$, the average energy gain for particles colliding with moving magnetized clouds under Lorentz invariance violation scenario results
\begin{equation}
  \left<\frac{\Delta E}{E}\right> = \left[2 + \frac{2}{3\sqrt{1+ \delta_n E^n}} \right] V^2 \textrm{.}
\end{equation}
Notice again that if $\delta_n=0$, the original result from Fermi is obtained.

\subsection{UHECR energy spectrum}
\label{subsec:uherc}

Following the calculations done by Blandford \& Eichler~\cite{1978ApJ...221L..29B}, the Fokker-Planck formalism~\cite{risken1984fokker,a} allows us to write the number of particles $N$ at a given moment in time with momentum between $p$ and $p+\delta p$ as
\begin{equation}
    \frac{\partial}{\partial \Vec{p}} \left[-\left<\frac{\Delta \Vec{p}}{\Delta t}\right>N + \frac{1}{2}\frac{\partial}{\partial \Vec{p}}\left<\frac{\Delta \Vec{p}\Delta \Vec{p}}{\Delta t}\right>N \right] - \frac{N}{\tau_{scp}} = 0 \textrm{,}
\end{equation}
where $\tau_{scp}$ is the escape time. The terms for the diffusion in space and time evolution were neglected because we are interested in steady-state sources. Re-writing this equation into energy space, we have the number of particles $N$ with energy between $E$ and $E+\delta E$ given by
\begin{multline}
   0 = \left[\frac{1}{2}\left( \frac{\partial E}{\partial p}  \right)^2 \Gamma \right] \frac{d^2 N}{dE^2} + \\
     + \left[\frac{\partial E}{\partial p} b + \frac{1}{2}\frac{\partial E}{\partial p} \frac{\partial}{\partial E}\left( \frac{\partial E}{\partial p}\right) \Gamma + \left(\frac{\partial E}{\partial p}\right)^2  \right] \frac{dN}{dE} + \\
    + \left[ \left( \frac{\partial E}{\partial p} \right)\frac{db}{dE} + \frac{1}{2}\frac{\partial E}{\partial p} \frac{\partial}{\partial E}\left( \frac{\partial E}{\partial p}\right) \frac{d\Gamma}{dE} \right.  +  \\
    + \left. \frac{1}{2}\left(\frac{\partial E}{\partial p}\right)^2 \frac{d^2\Gamma}{dE^2} - \frac{1}{\tau_{scp}} \right]
    N \textrm{,}
    \label{eq:en:spec:00}
\end{multline}
where we have used the definitions
\begin{equation}
  b \equiv - \frac{dE }{dt} \approx -\frac{\left<\Delta E\right>}{\tau_{avg}}
\end{equation}
and
\begin{equation}
  \Gamma \equiv \frac{d(\Delta E^2)}{dt} \approx \frac{\left<\Delta E^2\right>}{\tau_{avg}} \textrm{,}
\end{equation}
where $\tau_{avg}$ is the average time between collisions. We have also used
\begin{equation}
    \frac{\partial }{\partial p} = \left(\frac{\partial E}{\partial p} \right) \frac{\partial}{\partial E} \textrm{,}
\end{equation}
as well as the approximations of $\Delta p \approx \Delta E$ and the mass of the cloud being much larger than the mass of the particle, implying
\begin{equation}
    \left< \frac{\Delta p}{\Delta t} \right> = \frac{1}{2}\frac{\partial}{\partial p} \left< \frac{\Delta p \Delta p}{\Delta t} \right> \textrm{,}
\end{equation}
which, in our Lorentz-violating scenario, can be translated into
\begin{equation}
  b = - \frac{1}{2} \left(\frac{\partial E}{\partial p}\right) \frac{d\Gamma}{dE}
  \label{eq:b:01}
\end{equation}
and
\begin{equation}
  \frac{db}{dE} = \frac{1}{2} \left[\frac{\partial E}{\partial p}\frac{d^2\Gamma}{dE^2} + \frac{\partial}{\partial E}\left( \frac{\partial E}{\partial p}\right)\frac{d\Gamma}{dE} \right] \textrm{.}
  \label{eq:db:de}
\end{equation}

Using equations~\ref{eq:b:01} and~\ref{eq:db:de}, equation~\ref{eq:en:spec:00} becomes
\begin{multline}
    0 =  \left[\frac{1}{2}\left( \frac{\partial E}{\partial p}  \right)^2 \Gamma \right] \frac{d^2 N}{dE^2} + \\
    + \left[ \frac{1}{2}\frac{\partial E}{\partial p} \frac{\partial}{\partial E}\left( \frac{\partial E}{\partial p}\right) \Gamma + \frac{1}{2}\left(\frac{\partial E}{\partial p}\right)^2  \right] \frac{dN}{dE} - \left[ \frac{1}{\tau_{scp}} \right]N \textrm{.}
    \label{eq:en:spec:01}
\end{multline}

Using the energy dispersion relation modified by LIV (equation~\ref{eq:liv:edisper}), we can write
\begin{equation}
  \frac{\partial E}{\partial p} \approx \frac{1 + \left(\frac{2+n}{2}\right) \delta_n E^n}{\sqrt{1+\delta_n E^n}}
  \label{eq:de:dp}
\end{equation}
and
\begin{equation}
  \frac{\partial}{\partial E}\left( \frac{\partial E}{\partial p} \right) \approx \frac{\left(\frac{2+n}{2}\right)n \delta_n E^{n-1}}{1 + \left(\frac{2+n}{2}\right) \delta_n E^n} - \frac{n\delta_n E^{n-1}}{1+\delta_n E^n} \textrm{,}
  \label{eq:dde:de:dp}
\end{equation}
where the high-energy approximation $E \approx p$ was used again.

Inserting equations~\ref{eq:de:dp} and~\ref{eq:dde:de:dp} into equation~\ref{eq:en:spec:01}, leads to a second-order differential equation

\begin{equation}
  A(E,\delta_n,n)\frac{d^2N}{dE^2} + B(E,\delta_n,n)\frac{dN}{dE} + C(E,\delta_n,n)N = 0 \textrm{,}
  \label{eq:diff}
\end{equation}
such that our coefficients, after the previous calculations, are
\begin{equation}
    \frac{A(E, \delta_n,n)}{\alpha} = \frac{1}{2\alpha}\left[\frac{1 + \left(\frac{2+n}{2}\right) \delta_n E^n}{\sqrt{1+\delta_n E^n}}\right]^2 \Gamma \textrm{,}
\end{equation}
\begin{multline}
    \frac{B(E,\delta_n,n)}{\alpha} = \frac{1}{2\alpha}\left[ \frac{\left(\frac{2+n}{2}\right)n \delta_n E^{n-1}}{1 + \left(\frac{2+n}{2}\right) \delta_n E^n} - \frac{n\delta_n E^{n-1}}{1+\delta_n E^n} \right] \times \\
    \left[ \frac{1 + \left(\frac{2+n}{2}\right) \delta_n E^n}{\sqrt{1+\delta_n E^n}} \right]\Gamma + \\
    + \frac{1}{2\alpha}\left[\frac{1 + \left(\frac{2+n}{2}\right) \delta_n E^n}{\sqrt{1+\delta_n E^n}}\right]^2 \frac{d\Gamma}{dE} \textrm{,}
\end{multline}
\begin{equation}
    \frac{C(E,\delta_n,n)}{\alpha} = -\frac{1}{\alpha \tau_{scp}} \textrm{,}
\end{equation}
where $\alpha \equiv V^2/\tau_{avg}$.

Equation~\ref{eq:diff} can be numerically solved, as shown in Appendix~\ref{sec.app}. The solution is not straightforward and requires one extra assumption that the energy spectrum can be described by a power law in energy $N = N_0E^{\lambda (E)}$. The relation between the escape time and average time between collisions defines the index of the power law of the energy spectrum. We take $\frac{1}{\alpha \tau_{scp}} = 8/3$ to match the index of $-2$ as suggested by the UHECR data~\cite{PierreAuger:2021hun, ABBASI2023102864}.

Considering these assumptions, the solution for the energy spectrum is illustrated in Figures~\ref{fig:en:spec:2nd:order} and~\ref{fig:lambda:2nd:order}.

The index of the power law of the energy spectrum changes from $-2$ to approximately $-1$, for $n=1$, when the energy increases by two orders of magnitude and from $-2$ to approximately $-0.5$, for $n=2$, when the energy increases by one order of magnitude. Other values of $n$ entail different indexes, as shown in appendix \ref{sec.app}. Therefore, the introduction of LIV into the second-order Fermi mechanism results into an important modification of the energy spectrum of particles emitted by the source.

%time
Using Fermi's original argument, we can estimate the acceleration time under LIV assumption based on the mean free path $L$ between collisions
\begin{equation}
    t_{LIV}^{-1} = \frac{1}{2L} \frac{\left< \Delta E \right>}{E} \textrm{,}
    \label{eq:time}
\end{equation}
\begin{equation}
    t_{LIV} = \frac{L\left( 3\sqrt{1+\delta_n E^n} + 1\right)}{3\sqrt{1+\delta_n E^n}V^2} \textrm{.}
\end{equation}

Comparing our modified acceleration time ($t_{LIV}$) with the standard scenario ($t_{LI}$), we see that
\begin{equation}
    t_{LIV} = \frac{1 + 3\sqrt{1+\delta_n E^n}}{4\sqrt{1+\delta_n E^n}} t_{LI} \textrm{,}
\end{equation}
Note that $4\sqrt{1+\delta_n E^n}$ is always greater than $1 + 3\sqrt{1+\delta_n E^n}$, especially for ultra-high energies, implying that the time for particles to gain energy through this mechanism will decrease in the Lorentz Violation framework as we increase in energy.

%==================================================================
\section{First-order Fermi acceleration}
\label{sec:first:order}

Fermi's original argument was modified by introducing scattering of the particles in the shock waves. The works of Bell~\cite{10.1093/mnras/182.3.443}, Krymsk~\cite{1977DoSSR.234.1306K}, and Blandford \& Ostriker~\cite{BLANDFORD19871} laid the theoretical foundations of what would later be known as first-order Fermi mechanism. This mechanism relies on the propagation of a supersonic shock wave through the interstellar medium, where we assume the presence of a particle flux both in front of and behind the shock. As particles traverse the shock, they gain energy.

If a shock wave with velocity $U$ hits a particle with momentum $p_x$, both measured in the laboratory reference frame, and $V$  being the velocity of the downstream media in the particle reference frame, the energy of the particle in the laboratory reference frame after crossing the shock wave is
\begin{equation}
    E' = \gamma_V(E + p_x V) \textrm{.}
\end{equation}

It is possible to derive a relation between the upstream (1) and downstream (2) velocities in the laboratory frame, $v_1 = 4v_2$, supposing a fully ionized gas, leading to $V = \frac{3}{4}U$~\cite{Longair_2011,10.1093/mnras/182.3.443}.

Using equation~\ref{eq:liv:edisper}, we can write
\begin{equation}
    \frac{E' - E}{E} = \frac{\Delta E}{E} = \frac{V\cos{(\theta)}}{\sqrt{1+\delta_n E^n}} \textrm{.}
\end{equation}
Therefore, the average energy gain is
\begin{equation}
    \left< \frac{\Delta E}{E} \right> = \frac{4}{3\sqrt{1+\delta_n E^n}} V \textrm{.}
    \label{eq:liv:firstorder}
\end{equation}
If $\delta_n = 0$, we recover the LI result.
%================================
\subsection{UHECR energy spectrum}

Following Bell's argument~\cite{10.1093/mnras/182.3.443} with equation~\ref{eq:liv:firstorder}, the resulting energy spectrum can be written as
\begin{equation}
  \frac{dN}{dE} = -\left [  1 + \frac{2(1+\delta_n E^n) - n\delta_n E^n}{2(1+\delta_n E^n)^{3/2}}  \right] \frac{\sqrt{1+\delta_n E^n} N}{E} \textrm{.}
  \label{eq:espectrum:solution:1}
\end{equation}
If $\delta_n = 0$, we recover the LI result.

\begin{figure}[ht]
    \centering
    \includegraphics[width=0.8\textwidth]{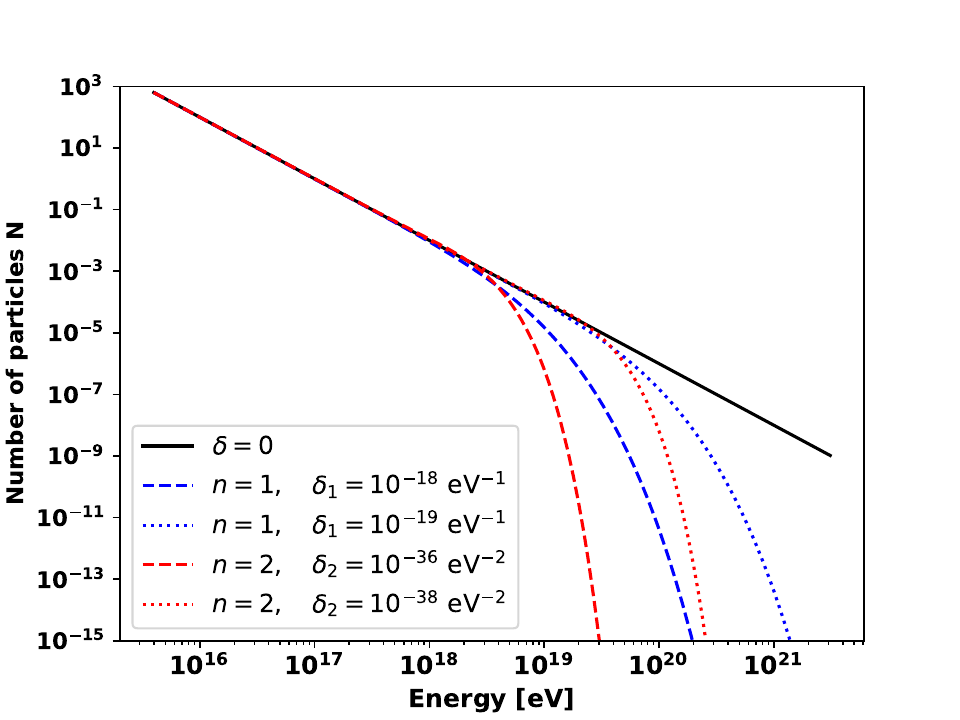}
    \caption{Energy spectrum of particles accelerated by first-order Fermi mechanisms with Lorentz Invariance Violation given by the $\delta$ parameter for the first $\delta_1$ and second $\delta_2$ orders of the energy dispersion relation expansion.}
    \label{fig:en:spec:1nd:order}
\end{figure}

\begin{figure}[ht]
    \centering
    \includegraphics[width=0.8\textwidth]{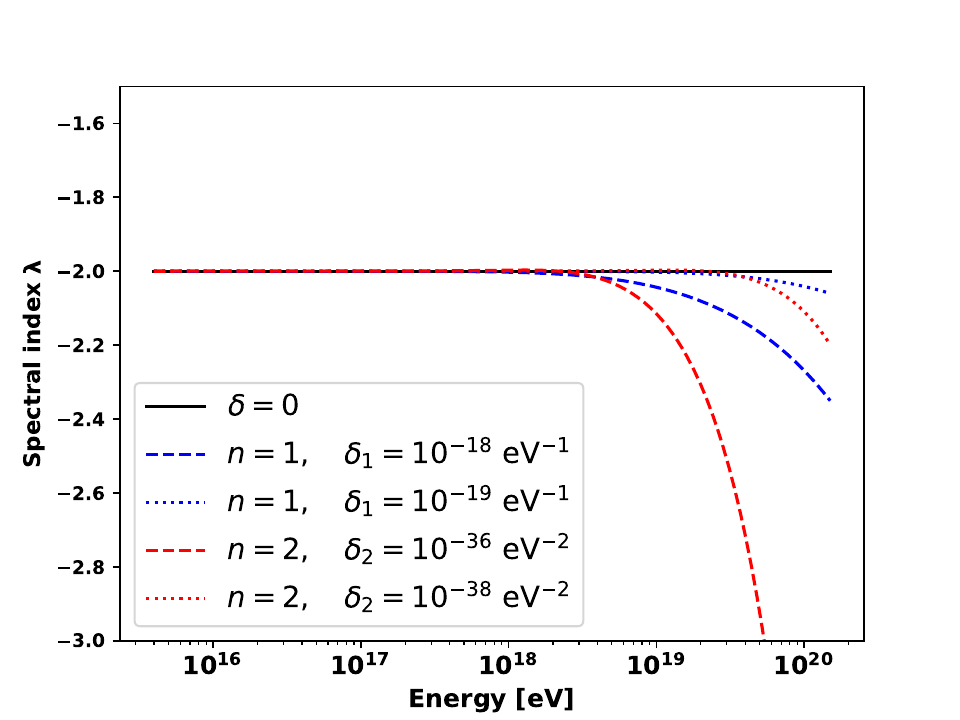}
    \caption{Spectral index of a power-law energy spectrum of particles accelerated by first-order Fermi mechanisms with Lorentz Invariance Violation given by the $\delta$ parameter for the first $\delta_1$ and second $\delta_2$ orders of the energy dispersion relation expansion.}
    \label{fig:lambda:1nd:order}
\end{figure}

Figure~\ref{fig:en:spec:1nd:order} shows the numerical solution of this equation. Figure~\ref{fig:lambda:1nd:order} shows the evolution of the power-law index of the energy spectrum.

The power-law index changes from $-2.0$ to $-3.0$ when the energy goes from $10^{18}$ to $5\times10^{19}$ eV for $n=2$ and $\delta_2=10^{-36}\ \textrm{eV}^{-2}$. Therefore, the introduction of LIV into the first-order Fermi mechanism results in an important modification of the energy spectrum emitted by the source.

\begin{figure}[ht]
    \centering
    \includegraphics[width=0.8\textwidth]{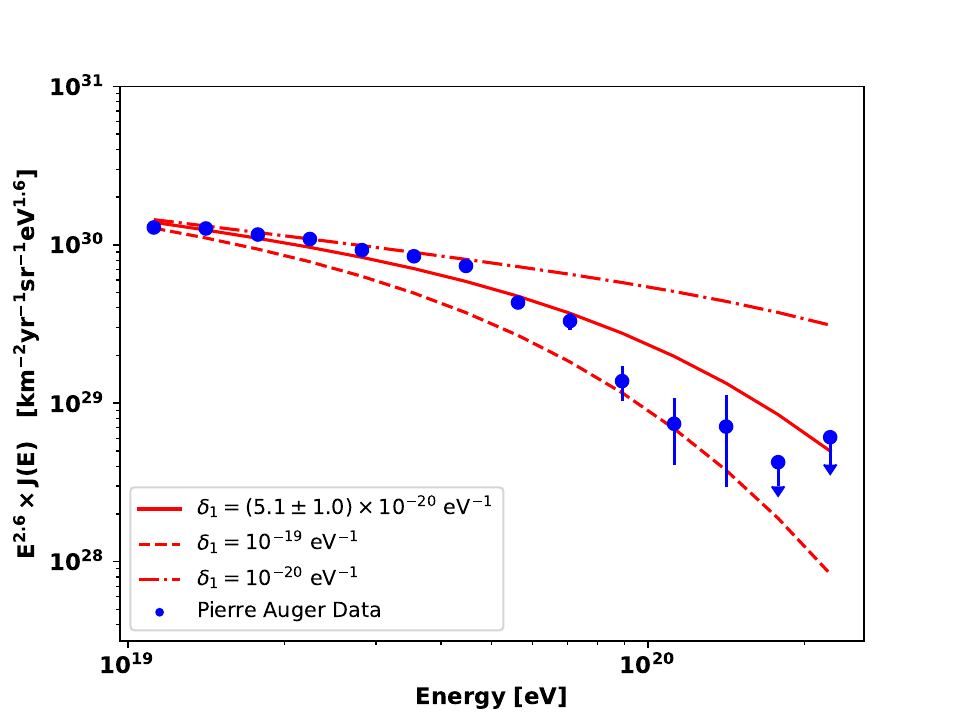}
    \caption{Energy spectrum of particles accelerated by first-order Fermi mechanisms with Lorentz Invariance Violation given by the $\delta$ parameter for the first $\delta_1$ order of the energy dispersion relation expansion. The blue dots show the energy spectrum measured by the Pierre Auger Observatory~\cite{PhysRevLett.125.121106}. The  $\delta_1$ parameter was fitted to the data. Different values of $\delta_1$ are shown. All spectrums have a normalization factor of $N_0 = \left(8.4 \pm 0.3 \right) \times 10^{41}$.}
    \label{fig:en:spec:1nd:order:auger:1}
\end{figure}

\begin{figure}[ht]
    \centering
    \includegraphics[width=0.8\textwidth]{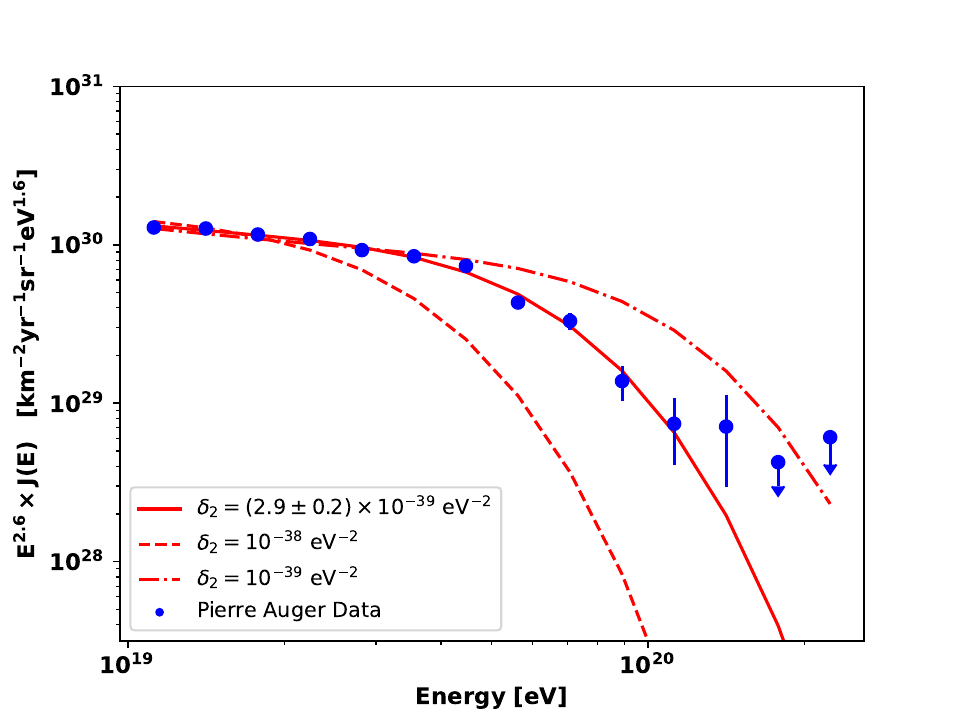}
    \caption{Energy spectrum of particles accelerated by first-order Fermi mechanisms with Lorentz Invariance Violation given by the $\delta$ parameter for the first $\delta_2$ order of the energy dispersion relation expansion. The blue dots show the energy spectrum measured by the Pierre Auger Observatory~\cite{PhysRevLett.125.121106}. The  $\delta_2$ parameter was fitted to the data. Different values of $\delta_2$ are shown. All spectrums have a normalization factor of $N_0 = \left(7.16 \pm 0.07\right) \times 10^{41}$.}
    \label{fig:en:spec:1nd:order:auger:2}
\end{figure}

The introduction of LIV in Fermi's first-order mechanism causes a suppression of the flux of particles for energies above the break, similar to that observed in measurements done by the Pierre Auger Observatory~\cite{PhysRevLett.125.121106}. Figures~\ref{fig:en:spec:1nd:order:auger:1} and~\ref{fig:en:spec:1nd:order:auger:2} show the energy spectrum measured by the Pierre Auger Observatory and the solution of equation \ref{eq:espectrum:solution:1} for different values of $\delta_1$ and $\delta_2$, respectvely. The plot is an illustration of the effect of LIV suppression in the Fermi first-order mechanism. It is not our intention to explain the data or limit LIV coefficients in this analysis. 

The maximum energy that an acceleration region can provide is of utmost importance; therefore, the effects of LIV must also be considered. For this system, we can write the acceleration time following the same definition as in equation \ref{eq:time}
\begin{equation}
    t = \frac{5}{3} \frac{D\sqrt{1+\delta_n E^n}}{V^2} \textrm{,}
\end{equation}
where $D$ is the Bohm diffusion coefficient~\cite{PhysRev.75.1851} for the best-case scenario, and we have used  $V = \frac{3}{4}U$ as above. Using the modified Larmour radius, $r_L = \frac{E\sqrt{1+\delta_n E^n}}{zeB}$, we obtained our new acceleration time
\begin{equation}
    t_{LIV} = \frac{5}{9}\frac{E(1+\delta_nE^n)}{zeBV^2} \textrm{.}
\end{equation}

Note that for the first-order mechanism, the acceleration time will now increase as we go to higher energies. From this, we can extract the maximum energy
\begin{equation}
    (1+\delta_nE_{max}^n)E_{max} = \frac{9}{5}zeB V^2 t_{LIV} \textrm{,}
\end{equation}
which, in realistic scenarios, considering current limits on the breaking parameters, will result in no significant deviations, achieving the Hillas limit~\cite{1984ARA&A..22..425H}.

%=============================================================
\section{Conclusion}
\label{sec:conclusion}

In this paper, we have calculated the acceleration of particles using the Fermi mechanisms of first and second order under the assumption of Lorentz invariance violation. LIV is introduced in a phenomenological approach by a modified dispersion relation. We calculated the resulting energy spectrum of particles accelerated in the source and the acceleration time.

For the second-order Fermi mechanism, the changes introduced by LIV entail a change in the spectral index for energies above the break. Such deviations depend on the order of the violation considered in equation \ref{eq:liv:edisper}. For the first-order mechanism, the changes introduced by LIV are important as the flux of particles is strongly suppressed for energies above the LI break, leading to a possible measurable effect. The suppression caused by LIV is of the same order of magnitude of the suppression measured in the data~\cite{PhysRevLett.125.121106}. The suppression is typically explained by the interaction of particles with background photons \cite{PhysRevLett.16.748} or by the maximum power of the sources \cite{1984ARA&A..22..425H,Aloisio_2023}. The addition of LIV in the acceleration mechanism could help complement these hypotheses without excluding them, as they are crucial effects at the highest energies.

The introduction of LIV in the two models leads to distinct changes. This difference can be explained by noting that, with the symmetry breaking, the acceleration time for the second-order mechanism tends to decrease compared to the standard case, facilitating the energy gain. However, for the first-order mechanism, the necessary time increases rapidly, resulting in the significant flux suppression of particles.

The calculation presented here can be used to derive limits on the LIV coefficients by using the measured UHECR energy spectrum. However, the calculation of the limits on the LIV coefficients depends on astrophysical assumptions such as the distribution of sources and the mass composition~\cite{Abreu_2022,Arsene_2021}. The potential of UHECR data to find or impose limits on LIV will improved significantly if a source is identified and if a subset of particles is selected. 

Most of the uncertainty in data analysis arises from the lack of knowledge regarding the type of particle arriving on Earth. The acceleration time, and consequently the maximum energy a particle can achieve at its source, is dependent on the particle's charge. Additionally, the total distance traveled by the particle from the source to Earth is influenced by its charge due to deviations caused by magnetic fields.

Furthermore, the ongoing challenge in identifying individual ultra-high-energy cosmic ray (UHECR) sources introduces significant uncertainty. Strong assumptions about source distributions and their characteristics further exacerbate this issue. Identifying a point source would eliminate many unknowns. Even if only the class of objects (e.g., AGN or starburst regions) responsible for these emissions is identified, it would allow for a deeper investigation into the details of the acceleration mechanisms involved.

For these reasons, the planned upgrades to the Pierre Auger Observatory~\cite{augerprime} and the Telescope Array Observatories~\cite{tax4} are expected to significantly enhance the detectability of Lorentz invariance violation (LIV) in the coming years. These upgrades aim to select a subset of proton events and identify local sources, thereby reducing the major uncertainties in the search for LIV.

The overall effect of LIV in the analysis of UHECRs must consider all three regions discussed in the introduction: the source, the extragalactic medium, and the atmospheric shower on Earth. It is implausible that Lorentz invariance is violated in only one of these regions and not the others. The calculations presented in this paper contribute to completing the overall scenario by incorporating considerations of the acceleration mechanism.

%======================================
%
\section*{Acknowledgments}

The authors are supported by the S\~{a}o Paulo Research Foundation (FAPESP) through grant number 2021/01089-1. VdS is supported by CNPq through grant number 308837/2023-1. MDF is supported by CAPES through grant number 88887.684414/2022-00. The authors acknowledge the National Laboratory for Scientific Computing (LNCC/MCTI,  Brazil) for providing HPC resources for the SDumont supercomputer (http://sdumont.lncc.br).

\bibliographystyle{plainnat}
\bibliography{references}

\appendix
\section{Appendix}
\label{sec.app}

\begin{figure}[ht]
    \centering
    \includegraphics[width=0.8\textwidth]{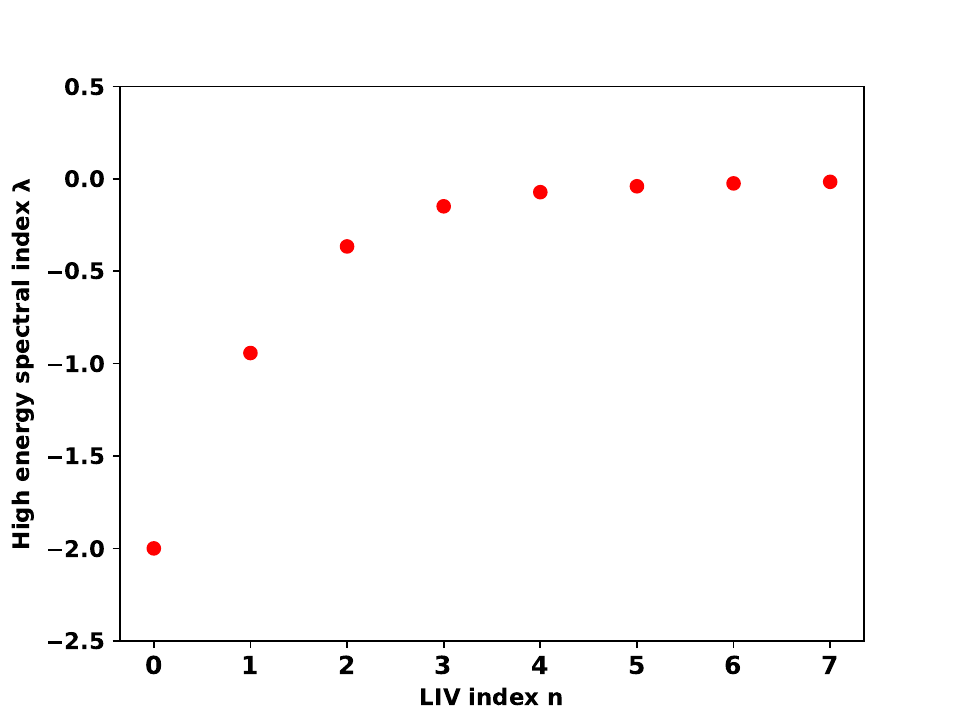}
    \caption{New power-law index for the second-order Fermi mechanism in the high-energy regime. The relation between characteristic times is taken to be $-8/3$. The $\delta$ parameter is fixed as it has no effect over the index result.}
    \label{fig:lambs}
\end{figure}

We present here the solution of equation

\begin{equation}
  A(E,\delta_n,n)\frac{d^2N}{dE^2} + B(E,\delta_n,n)\frac{dN}{dE} + C(E,\delta_n,n)N = 0 \textrm{,}
  \label{eq:dif:app}
\end{equation}
with coefficients given in section~\ref{subsec:uherc}. In the two limits, low ($\delta_n E^n <<1$) and high energy ($\delta_nE^n>>1$), the solutions can be envisaged. For the low energy regime, the LI result should be obtained; therefore, the solution is a power-law in energy with spectral index $-2$ when $1/\alpha \tau_{scp} = 8/3$ is assumed. For the high-energy regime, the Lorentz invariance violation becomes important, such that the equation will transform into
\begin{equation}
    \frac{(2+n)^2}{3}E^2\frac{d^2N}{dE^2} + \frac{(2+n)^2(2-n)}{3}E\frac{dN}{dE} - \frac{1}{\alpha \tau_{scp}} N = 0 \textrm{,}
    \label{eq:high}
\end{equation}
which is known as the Cauchy-Euler form~\cite{cauchy}\cite{euler} with solutions given by power laws with index determined by the coefficients. Figure~\ref{fig:lambs} shows the spectral indexes that solve this equation for different orders of LIV.
The solutions for low and high-energy regimes for all orders of $n$ are power laws. In each regime, the index of the power law is different. For the low-energy regime, the index is a constant determined by $\frac{1}{\alpha \tau_{scp}}$. The transition between the low and high-energy regimes can be modeled by assuming that the power-law index depends on the parameters of our problem
\begin{equation}
N(E,\delta_n,n) = N_0 E^{\lambda(E,\delta_n,n)}. 
\label{eq:sol:a}
\end{equation}

\begin{figure}[ht]
    \centering
    \includegraphics[width=0.75\textwidth]{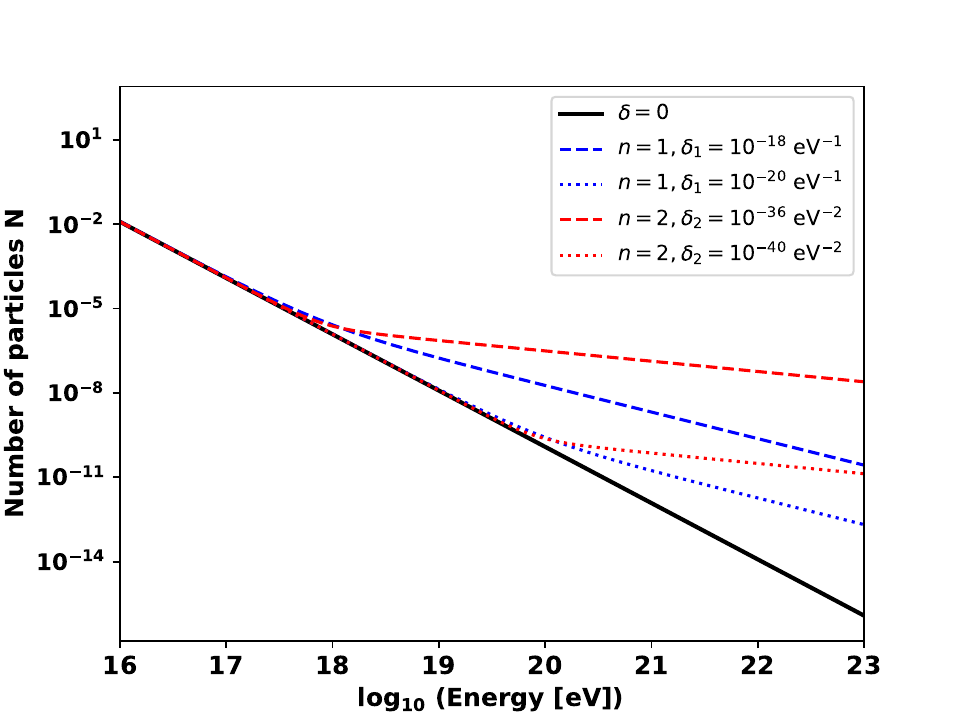}
    \caption{Energy spectrum of particles accelerated by second-order Fermi mechanisms with Lorentz Invariance Violation given by the $\delta$ parameter for first $\delta_1$ and second $\delta_2$ orders of the energy dispersion relation expansion.}
    \label{fig:en:spec:2nd:order}
\end{figure}

\begin{figure}[ht]
    \centering
    \includegraphics[width=0.75\textwidth]{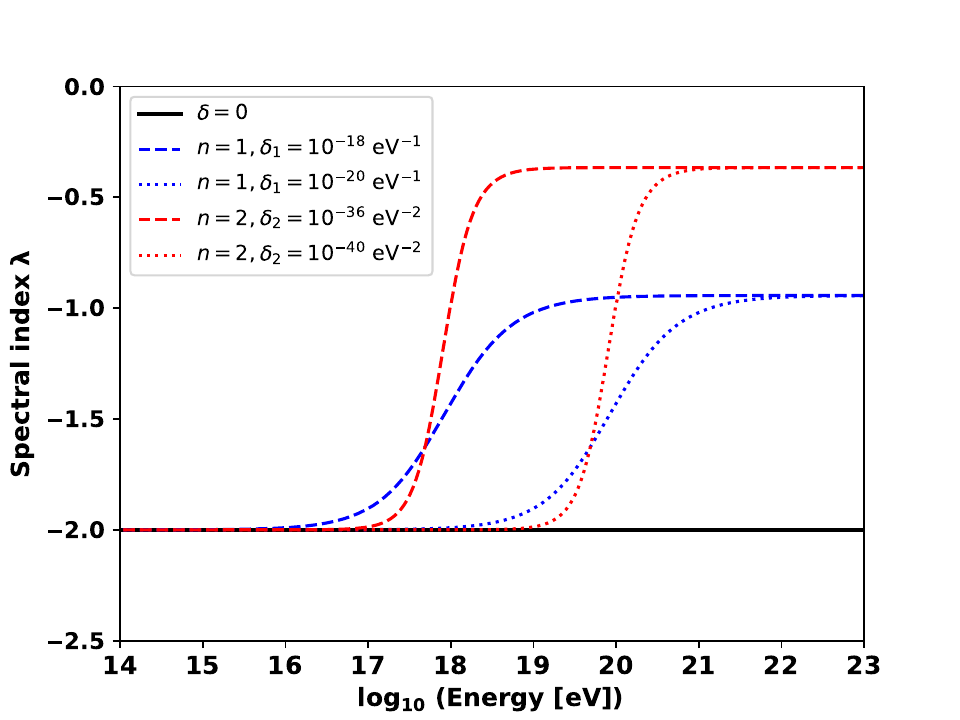}
    \caption{Spectral index of a power-law energy spectrum of particles accelerated by second-order Fermi mechanisms with Lorentz Invariance Violation given by the $\delta$ parameter for the first $\delta_1$ and second $\delta_2$ orders of the energy dispersion relation expansion.}
    \label{fig:lambda:2nd:order}
\end{figure}

Using this assumption in equation~\ref{eq:dif:app}, we can write
\begin{multline}
    A\left[ \left(\lambda' \ln{E} + \frac{\lambda}{E} \right)^2   +  \lambda'' \ln{E} + \left(\frac{2\lambda' E - \lambda}{E^2} \right) \right] + \\
    + B\left[ \lambda' \ln{E} + \frac{\lambda}{E} \right] + C = 0 \textrm{.}
    \label{eq:diff:lambda}
\end{multline}
The error function and $\tanh{(E)}$ can be shown to solve this equation. Both functions can describe the transition between low-energy (LI) and high-energy (LIV) regimes. These curves were normalized to match the spectral indexes derived from the analysis of equation \ref{eq:dif:app}. The solutions are discussed in section~\ref{subsec:uherc} and shown in Figures~\ref{fig:en:spec:2nd:order} and~\ref{fig:lambda:2nd:order}.

%===================================================

\end{document}